\documentclass[namedreferences]{SSRv}
\usepackage{epsfig}
\usepackage{graphicx}
\usepackage{times}
\usepackage{makeidx}

\begin{document}

\begin{article}
\begin{opening}

\title{Constraining fundamental constants of physics with \\
quasar absorption line systems}

\author{Patrick  \surname{Petitjean}$^{1}$,
	Raghunathan  \surname{Srianand}$^{2}$,
	Hum  \surname{Chand}$^{3}$,
	Alexander  \surname{Ivanchik}$^{4}$,
	Pasquier  \surname{Noterdaeme}$^{2}$,
        Neeraj  \surname{Gupta}$^{5}$
%      C. C.    \surname{CUTHOR}$^{4}$
        }

\runningauthor{Patrick Petitjean et al.}
\runningtitle{Quasar absorption lines}

\institute{$^{1}$ UPMC Paris 06, Institut d'Astrophysique de Paris,
          UMR7095 CNRS, 98bis Boulevard Arago, F-75014, Paris, France \email{petitjean@iap.fr}\\
          $^{2}$ IUCAA, Post Bag 4, Ganesh Khind, Pune 411 007, India \email{anand@iucaa.ernet.in}\\
          $^{3}$ ARIES, Manora Peak, Nainital-263129 (Uttarakhand), India \email{hum@aries.ernet.in}\\
          $^{4}$ Ioffe Physical Technical Institute, St Petersburg 194021, Russia \email{iav@astro.ioffe.ru}  \\             
          $^{5}$ Australia Telescope National Facility, CSIRO, Epping, NSW 1710, Australia \email{Neeraj.Gupta@atnf.csiro.au}\\
}

\date{Received ; accepted }

\begin{abstract}
We summarize the attempts by our group and others to derive constraints on variations of fundamental
constants over cosmic time using quasar absorption lines. Most upper limits reside in the 
range 0.5$-$1.5$\times$10$^{-5}$ at the 3$\sigma$ level over a redshift range of approximately
$0.5 - 2.5$ for the fine-structure constant, $\alpha$, the proton-to-electron mass ratio, $\mu$
and a combination of the proton gyromagnetic factor and the two previous constants, 
$g_{\rm p}(\alpha^2/\mu)^{\nu}$, for only one claimed variation of $\alpha$. It is therefore very important to  
perform new measurements to improve the sensitivity of the numerous methods to at least $<$0.1$\times$10$^{-5}$ 
which should be possible in the next few years. Future instrumentations on ELTs in the optical
and/or ALMA, EVLA and SKA pathfinders in the radio will undoutedly boost this
field by allowing to reach much better signal-to-noise ratios at higher spectral resolution
and to perform measurements on molecules  in the ISM of high redshift galaxies.
%The strongest constraints may come from radio observations of molecules in the ISM of high redshift galaxies.
\end{abstract}

\end{opening}
 \vspace{-0,5cm}
{\small

\section{Introduction}\label{Introduction}

\noindent As most of the successful physical theories rely on the constancy of
few fundamental quantities (the speed of light, $c$,
the fine-structure constant, $\alpha$, the proton-to-electron mass ratio,
$\mu$, etc), constraining the possible time variations of these fundamental
physical quantities is an important step toward understanding the rules of nature.
Current laboratory constraints exclude any significant time variation of the dimensionless
constants in the low-energy regime. It is not excluded however that they could have 
varied over cosmological time-scales. Savedoff (1956) first 
pointed out the possibility of using redshifted atomic lines from distant objects 
to test the evolution of dimensionless physical constants. 
The idea is to compare the wavelengths of the same transitions measured in the 
laboratory on earth and in the remote universe.  
This basic principle has been first applied to QSO absorption lines by Bahcall et al. (1967). 
The field has been given tremendous interest recently with the advent of 
10~m class telescopes. 
\par\noindent
For comparison with constraints obtained from laboratory experiments see
other papers in this volume and reviews by others e.g. Uzan (2003) or Flambaum (2008).
\\
%\cite[YYYY]{alias} 
\section{The method}\label{QSOals}
The idea is simply to compare wavelengths of the same transition measured in the remote 
universe and in the laboratory. As we live in an expanding universe we need at least
two transitions that have different sensitivities to the changes in fundamental
constants. Ideally, transitions with no sensitivity to a variation of constants
are used to measure the redshift and transitions with a large sensitivity
to a variation of constants are used to constrain this variation once
the redshift is known.
\subsection{Atomic data}
Modern spectrographs mounted on 10~m class telescopes provide high signal-to-noise
ratio data on faint remote quasars at very high spectral resolution (typically $R$~$\sim$~50,000).
Since for a given spectral resolution, rest-frame measurements at high redshift have a better
precision by a factor of 3 to 5 than laboratory determinations (see e.g. Petitjean \& Aracil 2004), 
the method often requires dramatic improvements in laboratory measurements (see references in Murphy et al.
2003 for $\alpha$, Ubachs \& Reinhold 2004, Philip et al. 2004, Ivanov et al. 2008 for $\mu$). 

Atomic calculations are also needed to determine the sensitivity coefficients that
characterize the change in rest wavelength due to a change in a given constant. 
This has been done for $\mu$ from $H_2$ lines (see Varshalovich \& Potekhin 1995)
and from molecular lines (e.g. Flambaum \& Kozlov 2007)
and for $\alpha$ (e.g. Dzuba et al. 2002 and references therein; Kozlov et al. 2008a,b)

\subsection{Quasar absorption line systems}
\vskip -0.2truecm
The measurements are performed using absorption systems seen in the spectra
of remote quasars. Fig.~\ref{QSOspectrum} shows a quasar spectrum  obtained
after a typical observation of 10 hours with UVES at the European Very Large Telescope. The quasar,
at a redshift of $z_{\rm em}$~=~2.58, is quite bright and can be observed at high 
spectral resolution. Its spectrum is characterized by emission lines from the 
Lyman series of neutral hydrogen (Lyman-$\alpha$, normally at 1215\AA~  is redshifted 
by a factor 1~+~$z_{\rm em}$~=~3.58 at $\sim$4340~\AA) or the resonance transitions 
from C$^{3+}$ (or C~{\sc iv}) normally at $\sim$1550~\AA~ and seen here at $\sim$5550~\AA. 
It can be seen that numerous absorption lines are superimposed on top of the continuum
from the quasar. These absorptions arise when the line of sight crosses
by chance a gaseous cloud. At wavelengths smaller than the Lyman-$\alpha$
emission line from the quasar, any tiny amount of neutral hydrogen (corresponding to
intergalactic clouds) will produce a narrow Lyman-$\alpha$ absorption line. 
This region is called the Lyman-$\alpha$ forest. Measurements try to avoid 
this region of the spectrum as any absorption can be blended with an intervening 
H~{\sc i} Lyman-$\alpha$ line. This is not possible however for H$_2$ which UV lines are
always redshifted in the Lyman-$\alpha$ forest. When the line of sight passes through 
the halo or disk of a galaxy, a strong Lyman-$\alpha$ absorption is seen together
with metal lines. The strongest Lyman-$\alpha$ lines (with column densities log~$N$(H~{\sc i})~$>$~20.3) 
correspond to the so-called damped Lyman-$\alpha$ systems (see Fig.~1).  

\begin{figure}
%\centerline{\includegraphics[width=9cm]{fig-name.pdf}}
%\centerline{\includegraphics[width=9cm]{SpectreQSO.jpg}} 
%\centerline{\includegraphics[width=9cm]{PetitjeanF1.jpg}}
\centerline{\includegraphics[width=10.5cm,angle=270]{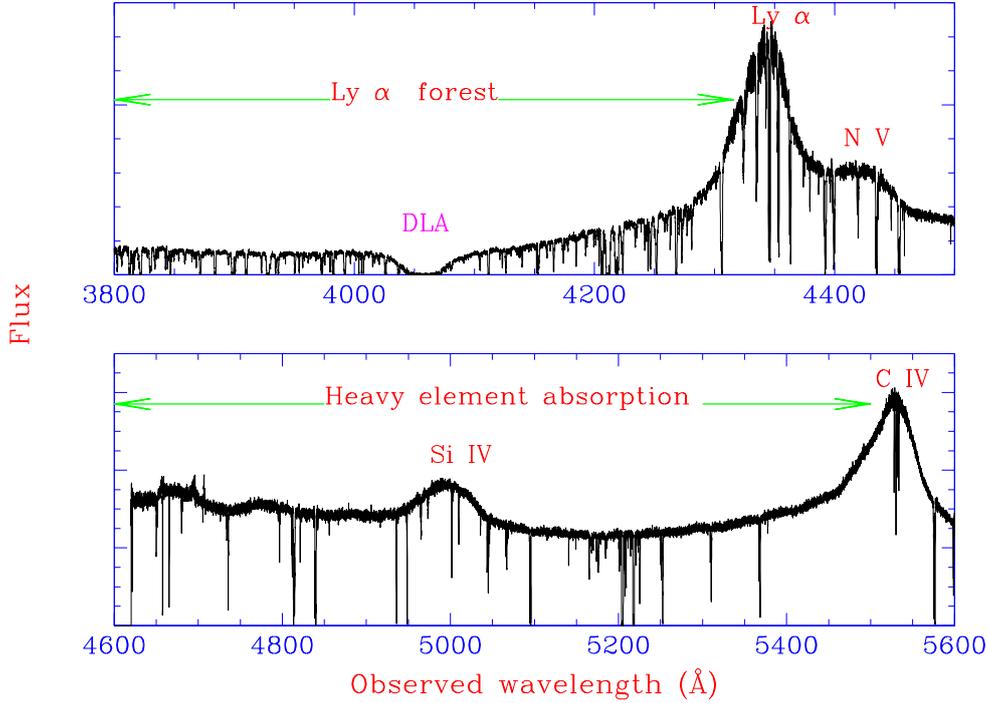}}
\caption{UVES-VLT spectrum of a quasar with emission redshift $z_{\rm em} = 2.58$.
The quasar is characterised by broad emission (H~{\sc i} Lyman-$\alpha$ at 
$\lambda\sim 4340$~\AA, or C~{\sc iv}$\lambda$1550 at $\lambda\sim 4340$~\AA). 
Below 4350~\AA, numerous H~{\sc i} Lyman-$\alpha$ absorption lines are seen that are produced by  
intergalactic clouds (narrow lines) or galactic disks (the so-called "Damped Lyman-$\alpha$
systems or DLAs) located by chance at smaller redshift along the line of sight to the quasar.
A DLA system is present along this line of sight at $z_{\rm abs}$~=~2.33 (Lyman-$\alpha$
absorption at $\sim$4050~\AA.) Metal lines are seen above 4350~\AA.
}
\label{QSOspectrum}
\end{figure} 
%\subsection{The method}
Small variations in the constants induce small positive or negative shifts in
the wavelengths of atomic or molecular species. It must be realized that these
shifts are quite small. For a relative variation of $\sim$10$^{-5}$
in the fine-structure constant $\alpha$, the typical shift of transitions
easily observable is $\sim$0.5~km/s (the situation is slighly better in the radio,
see below). This means about 20~m\AA~ for a redshift of about $z\sim2$.
This corresponds to about a third of a pixel at the spectral resolution of $R\sim 40000$ achieved
with UVES-VLT or HIRES-Keck. This kind of measurement is not
easy because (i) the number of independent absorption lines used in an individual
measurement is not large, typically five or six, and (ii) several sources of uncertainties 
hamper the measurement.

The dependence of rest wavelengths to the variation  of $\alpha$ 
is  parameterized using the fitting function  given by Dzuba et al. (1999), 
%\begin{equation} 
$\omega=\omega_{0}+q_{1}x+q_{2}y$. 
%\end{equation} 
Here $\omega_{0}$ and $\omega$ are, respectively, the vacuum wave  
number (in  units of cm$^{-1}$) measured in the laboratory  
and in the absorption system at  redshift $z$. $x$ and $y$ are dimensionless 
numbers defined as $ x=(\alpha_{z}/\alpha_{0})^{2}-1$ and 
$ y=(\alpha_{z}/\alpha_{0})^{4}-1 $. The  sensitivity  
coefficients $q_{1}$ and $q_{2}$ are obtained using  
many-body relativistic calculations (see Dzuba et al. 1999). 

\subsection{Source of errors}
\vskip -0.2truecm
The absorption lines have complex profiles because they are the result
of the QSO photons travelling through the highly inhomogeneous medium that
is associated with the potential wells of cosmological halos. 
These profiles are usually fitted using a combination of Voigt-profiles.
For each component, the exact redshift, the column density and the 
width of the line (Doppler parameter) are fit parameters to be determined
in addition to the shift from a possible variation of constants.
These parameters are constrained assuming that the profiles are the same for
all transitions. This is obviously true for transitions from the 
same species (as in Fig.~\ref{fit}) but is not necessarily true
in case transitions from different species are used for the same
measurement. Indeed, for testing the variations of $\alpha$, transitions from Mg~{\sc ii},
Si~{\sc ii} and/or Fe~{\sc ii} are commonly used. To avoid this problem, one could use transitions from
one species only (Quast et al. 2004, Chand et al. 2005, 
Levshakov et al. 2006) but suitable systems are rare and
the sensitivity of the method is reduced.

Another difficulty is that the fit is usually not unique. This is not a severe
problem if the lines are not strongly saturated however (as in the
bottom panel of Fig.~\ref{fit}) because in that case the positions of
the components are well defined. Simulations have shown (Chand et al. 2004)
that the presence of strongly saturated lines can increase errors
on the determination of $\Delta\alpha$/$\alpha$ by a factor of two in 
the case of a simple profile.

\begin{figure}
%\centerline{\includegraphics[width=9cm]{fig-name.pdf}}
%\centerline{\includegraphics[width=9cm]{example_si4fit.jpeg}} 
\centerline{\includegraphics[bb=16 140 579 724,width=10cm,clip=true]{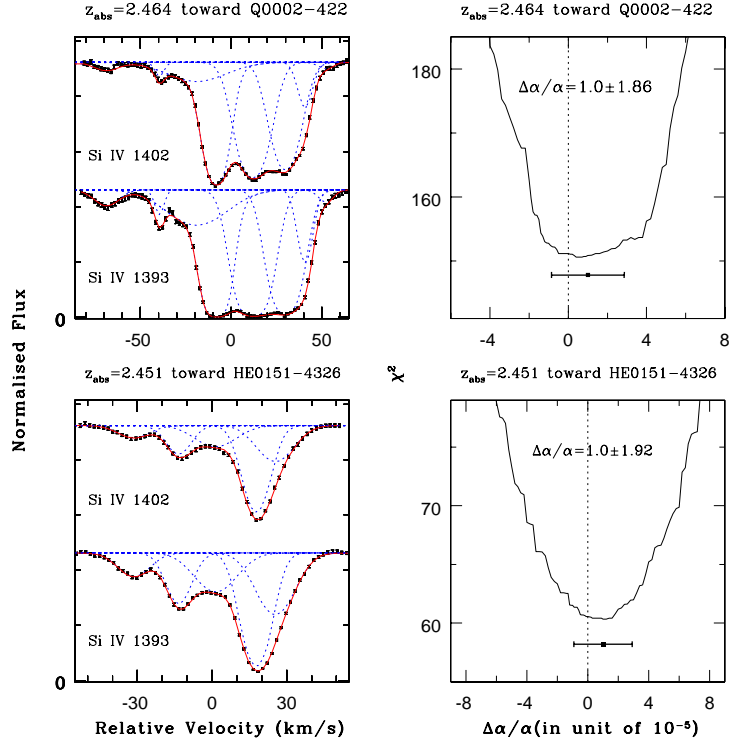}} 
\caption{Fits of Si~{\sc iv} doublets. Absorption
profiles are decomposed in Voigt-profile components (shown as dotted lines). 
Errors are increased in case the profile is strongly saturated.
$\chi^2$ curves are shown in the right-hand panels.
Note that the SNR is particularly good in the present cases.}
\label{fit}
\end{figure}

Spectra in the optical are taken with high resolution echelle spectrographs that have usually
a red and a blue arm. In each arm the spectrum is split into a large
number of orders and recorded on different CCDs. The wavelength calibration solution
is calculated for each CCD. This can introduce {\sl local} deviations from
the correct solution that may be difficult to control. Wavelength calibration
is of high importance here and should be carefully controlled (see Thompson et al. 2008
in the case of UVES). A way to overcome the difficulty is to use completely independent 
instruments to observe the same object. Chand et al. (2005) used the very stable spectrograph 
HARPS mounted on the 3.6~m telescope at the La~Silla observatory to observe the bright 
quasar HE~0515$-$4414 that was previously observed with UVES. They find that although the 
dispersion of the wavelength solution is much smaller for HARPS than for
UVES, errors are well constrained except for {\sl local} artifacts.
This means that eventhough wavelength calibration is not
a concern at the sensitivity level we can expect to achieve, local
problems can spoil the measurement for a few systems.
Also for some methods, two absorptions lines in very different 
wavelength bands (e.g. the radio and the UV) are used (see Section 5.1) and
the intercalibration has to be checked carefully.

Note that temperature must be controlled precisely or registered carefully
so that adequate air-vacuum correction can be applied. Flexures in the instrument are 
dealt with by recording a calibration lamp spectrum before and after the
science exposure which is usually 1~hour long. 
The signal-to-noise ratio of the data is a crucial issue.
Simulations by Chand et al. (2004) showed that errors are inversely correlated
with SNR. A minimum of SNR~$\sim$~50 at the position of the absorption lines
is required to achieve measurements at the level considered here.
%of the variation of $\alpha$ at the 10$^{-5}$ precision level.

%
\section{Variation of the fine-structure constant $\alpha$}\label{Sec1}
\noindent 
\vskip -0.2truecm
\subsection{The Many-Multiplet Method}\label{}
\vskip -0.2truecm
The power of the Many-Multiplet Method (MMM) is to use a large number of transitions
to constrain the variation of $\alpha$. At least five transitions are
used, usually from different species. The transitions are choosen
so that their sensitivities to a change in $\alpha$ are different.
For example rest wavelengths of Mg~{\sc ii} doublets 
and Mg~{\sc i} are fairly insensitive to small changes in 
$\alpha$ thereby providing good anchors for measuring the 
systemic redshift. Whereas the rest wavelengths 
of Fe~{\sc ii} multiplets are very sensitive to small variations 
in $\alpha$. The  accuracy depends on how well the absorption line 
profiles are modeled. 
The recent application of the many-multiplet method (Dzuba et al. 1999, Webb et al. 1999) has 
improved by an order of magnitude the accuracy of the $\Delta\alpha/\alpha$~ 
measurements based on QSO absorption lines (Webb et al. 2001). 
Analysis of HIRES/Keck data has resulted in the claim of a variation in  
$\alpha$, {$\Delta\alpha/\alpha$~ = ${(-0.54\pm0.12)}\times10^{-5}$}, over a 
redshift range  0.2~$<$~$z$~$<$~3.7 (Murphy et al. 2003). 

In order to check this result independently, we have applied the MM method 
to very high quality (SNR$\sim60-80$, R$\ge 44,000$) UVES/VLT data.
In view of the numerous systematic errors involved in the MM
method, we have carried out detailed simulations to define proper
selection criteria to choose suitable absorption systems in order
to perform the best analysis (see Chand et al. 2004 for details). 
Application of these selection criteria to the full sample  of 50 Mg~{\sc ii}/Fe~{\sc ii} 
systems lead us to restrict the study to 23 Mg~{\sc ii}/Fe~{\sc ii} systems  
over a redshift range $0.4\le z\le 2.3$.
The weighted mean of the individual measurements from this analysis is
a non detecton with a 3$\sigma$ upper limit of 
${\Delta\alpha/\alpha}$~$<$~${0.20\times10^{-5}}$ (Srianand et al. 2004, Chand et al. 2004).

All further analysis performed with UVES spectra fail to confirm any 
variation in $\alpha$ (Quast et al. 2004; Levshakov et al. 2005). 
In particular, Chand et al. (2006) analyse spectra of the bright quasar
HE 0515$-$4414 taken with two different instruments, UVES at the VLT
and HARPS at the 3.6~m telescope in La~Silla. They show that the results of 
a non-evolving $\alpha$ reported in the literature based on UVES/VLT data 
should not be heavily influenced by problems related to wavelength calibration 
uncertainties and multiple component Voigt profile decomposition. Considering 
that different procedures can be used, a robust 3$\sigma$ limit on the variation of $\alpha$
at $z\sim 1.5$ obtained with UVES data is ${\Delta\alpha/\alpha}$~$<$~0.30$\times$10$^{-5}$.
 
Note that several authors have used the five transitions of Fe~{\sc ii}
to obtained a limit from absorption lines of only one species in order to be
certain that the profile structures are identical (Quast et al. 2004, 
Chand et al. 2006, Levshakov et al. 2007). Useful atomic data are given by Porsev et al. (2007).
The number of systems suitable for such measurements is unfortunately very 
small and limits achieved are of the same order of magnitude.

It is now of high importance to improve the procedure and to increase the number of
measurements in order to decrease this limit to below 10$^{-6}$ which is 
a reasonable goal for present day instrumentation. 

%\subsubsection{}\label{}
\subsection{The alkali doublet (AD) method}\label{}
\vskip -0.2truecm
Alkali doublets are conspicuous in astrophysical spectra both in
emission (for example the [O~{\sc iii}]$\lambda\lambda$4969,5007 doublet) and
in absorption (for example the Si~{\sc iv}$\lambda\lambda$1393,1402
or the C~{\sc iv}$\lambda\lambda$1548,1550 doublets). In the later case however
atomic data are not well known (e.g. Petitjean \& Aracil 2004).
The method, although less sensitive than the MM method, has the advantage to
use only one species and to be applicable to higher redshifts.
Bahcall et al. (1967) were the first of a long list to  apply this technique to
QSO spectra.
%. Their analysis provided 
%${\Delta\alpha/\alpha} \equiv(\alpha_{z}-\alpha_{0})/\alpha_{0}= (-2\pm 5)\times 10^{-2}$ 
%at a redshift $z$~$\sim$~1.95. 
%Since then several authors have used  the AD method to constrain the variation of $\alpha$ 
%(Wolfe et al. 1976, Levshakov 1994, Potekhin \& Varshalovich 1994, 
%Cowie \& Songaila 1995, Varshalovich et al. 1996, Martinez et al. 2003).  

More recently, Murphy et al. (2001) analysed a KECK/HIRES sample of 21 Si~{\sc iv} doublets  
observed along 8 QSO sight lines and derived $\Delta\alpha/\alpha$~$<$~${3.9\times10^{-5}}$.
The analysis of 15 Si~{\sc iv} doublets selected from a ESO-UVES sample yielded 
the strongest constraint obtained with this method: $\Delta\alpha/\alpha < 1.3\times10^{-5}$ over  
the redshift range ${ 1.59\le z\le 2.92}$ (Chand et al. 2005).

The AD method can be applied to emission as well as absorption lines.  
However emission lines are usually broad as compared to absorption lines. Errors are therefore
larger on individual measurements and must be beaten by large statistics. 
As a result, the constraints obtained from emission lines are not as strong as 
those derived from  absorption lines. Bahcall et al. (2004) have recently found 
$\Delta\alpha/\alpha < 4.2\times10^{-4}$ using O~{\sc iii} 
emission lines from SDSS QSOs.  
%
%\paragraph{ .}

%\begin{figure}%
%\centerline{\includegraphics[width=9cm]{fig-name.pdf}}
%\centerline{\includegraphics[width=9cm]{fig-name.eps}} 
%\caption{ 
%.}\label{fig-name}
%\end{figure}

%\begin{equation}\label{Eq-name}
% .
%\end{equation}
\section{Variation of the proton-to-electron mass ratio $\mu$}\label{Sec2}
\noindent \vskip -0.2truecm
In the framework of unified theories (e.g. SUSY GUT) with a common
origin of the gauge fields, variations of the gauge coupling
$\alpha_{\rm GUT}$ at the unified scale ($\sim10^{16}$~GeV) will
induce variations of all the gauge couplings in the low energy
limit, $\alpha_{\rm i}=f_{\rm i}(\alpha_{\rm GUT},E)$, and provide a relation
$\Delta\mu/\mu\simeq R \Delta\alpha/\alpha$, where $R$ is a model
dependent parameter and $|R|\le 50$ (e.g. Dine et al. 2003,
and references therein). Thus, independent
estimates of $\Delta \alpha/\alpha$ and $\Delta \mu / \mu$ could
constrain the mass formation mechanisms in the context of unified
theories.

On earth, the proton-to-electron mass ratio has been measured
with a relative accuracy of $2\times 10^{-9}$ and equals $\mu_0 =
1836.15267261(85)$. Laboratory
metrological measurements rule out considerable variation of $\mu$
on a short time scale but do not exclude its changes over the
cosmological scale, $\sim 10^{10}$ years. Moreover, one can not
reject the possibility that $\mu$ (as well as other constants)
could be different in widely separated regions of the Universe.

\subsection{H$_2$}
The method using H$_2$ transitions to constrain the possible variations of $\mu$
was proposed by Varshalovich and Levshakov (1993). It
is based on the fact that wavelengths of electron-vibro-rotational
lines depend on the reduced mass of the molecule, with the
dependence being different for different transitions. It enables
us to distinguish the cosmological redshift of a line from the
shift caused by a possible variation of $\mu$.

Thus, the measured wavelength $\lambda_{\rm i}$ of a line formed in the
absorption system at the redshift $z_{\rm abs}$ can be written as,
%\begin{equation}
$\lambda_{\rm i}=\lambda_{\rm i}^0(1+z_{\rm abs})(1+K_{\rm i} \Delta\mu/\mu)$,
%\label{era}
%\end{equation}
where $\lambda_{\rm i}^0$ is the laboratory (vacuum) wavelength of the
transition, and $K_{\rm i}= \mbox{d} \ln \lambda_i^0/ \mbox{d} \ln \mu$
is the sensitivity coefficient for the Lyman and Werner
bands of molecular hydrogen (Varshalovich and Potekhin 1995). This expression 
can be represented in terms of the individual line redshift 
$z_{\rm i} \equiv \lambda_{\rm i}/\lambda_{\rm i}^0-1$ as,
%\begin{equation}
$z_{\rm i}=z_{\rm abs}+b K_{\rm i}$, 
%\label{korr}
%\end{equation}
where $b=(1+z_{\rm abs})\Delta\mu/\mu$. 
%Note, in case of nonzero
%$\Delta\mu/\mu$, $z_{\rm abs}$ do not equal to the standard mean
%value which is $\overline{z}= (\Sigma \, z_{\rm i})/N$.

In reality, $z_{\rm i}$ is measured with some uncertainty which is
caused by statistical errors of the astronomical measurements,
$\lambda_{\rm i}$, and by errors of the laboratory measurements of
$\lambda_{\rm i}^0$. Nevertheless, if $\Delta \mu / \mu$ is nonzero,
there must be a correlation between $z_{\rm i}$ and $K_{\rm i}$ values.
Thus, a linear regression analysis of these quantities yields $z_{\rm abs}$
and $b$ (as well as its statistical significance), consequently an
estimate of $\Delta\mu/\mu$.

Several studies have yielded tight upper limits on
$\mu$-variations, $|\Delta\mu/\mu|<7\times10^{-4}$ (Cowie \& Songaila 1995),
$|\Delta\mu/\mu|<2\times10^{-4}$ (Potekhin et al. 1998),
$|\Delta\mu/\mu|<5.7\times10^{-5}$ (Levshakov et al. 2002) and 
$\Delta\mu/\mu<7\times10^{-5}$ (Ivanchik et al. 2003).
Recently, a new limit was estimated, $\Delta\mu$/$\mu$$<$2.2$\times$10$^{-5}$ at the 3$\sigma$ level,
by measuring wavelengths of 76 H$_2$ lines of Lyman and Werner bands 
from two absorption systems at $z_{\rm abs}=2.5947$ and
$3.0249$ in the spectra of quasars Q~0405$-$443 and Q~0347$-$383, respectively.
Data were of the highest spectral resolution ($R$~=~53000) and S/N ratio (30$-$70)
for these kind of studies (Ivanchik et al. 2005). 

This result is subject to important systematic errors of two kinds: (i)
using different sets of laboratory wavelengths yield different results; (ii) the 
molecular lines are located in the Lyman-$\alpha$ forest where they can be
strongly blended with intervening H~{\sc i} Lyman-$\alpha$ absorption lines.
The first type of systematics are addressed by new laboratory measurements (Philip et al. 2005;
Reinhold et al. 2006). The second type of systematics needs careful fitting of the Lyman-$\alpha$
forest. This has been performed recently by King et al. (2008). These authors use principally the
same set of data as above and derive  $\Delta\mu$/$\mu$$<$1.2.$\times$10$^{-5}$ at the 
3$\sigma$ level. 

\subsection{HD}
\vskip -0.2truecm
The detection of several HD transitions makes it possible to test the 
possible time variation of the proton-to-electron mass ratio, in the same way as
with H$_2$ but in a completely independent way.  
As these measurements may involve various unknown systematics, it is important to use different sets of lines
and different techniques. Sensitivity coefficients and accurate wavelengths for HD transitions 
have been published very recently (Ivanov et al. 2008). 
It must be noted however that till now H$_2$ is detected in absorption in only 14 Damped Lyman-$\alpha$ systems 
whereas HD is detected in only two places in the whole universe. 
%\begin{figure}
%\includegraphics[width=\hsize,bb=23 410 590 750]{fig4.ps}
%\caption{\label{mu} The relative position of the lines 
%$\zeta_i$ is plotted as a function of the sensitivity coefficient $K_i$. 
%The three error bars at each $K$ correspond to individual components
%that are slightly shifted along the x-axis for clarity. 
%Filled circles represent the weighted mean for each transition. The
%slope of the linear fit gives directly $\Delta \mu/\mu$, while
%the associated 95\% confidence interval is delimited by the two curves. 
%}
%\end{figure}
\par\noindent
Deuterated molecular hydrogen was detected very recently together with carbon monoxide (CO; Srianand et al. 2008)
and H$_2$ in a Damped Lyman-$\alpha$ cloud at $z_{\rm abs}=2.418$ toward the quasar SDSS1439$+$11. 
Five lines of HD in three components were detected together with more than a hundred
H$_2$ transitions in seven components. Although each HD component is associated to one of the H$_2$ components, 
the strong blending of the latter, especially in low rotational levels, does not allow for the exact
determination of 
%$N($HD$)/2N($H$_2)$ in individual components and/or 
the relative positions of the HD and H$_2$ components. 
In passing, the column densities integrated over the whole 
profile for both HD and H$_2$ yield $N({\rm HD})/2N({\rm H_2})=1.5^{+0.6}_{-0.4}\times10^{-5}$
(Noterdaeme et al. 2008).
Five HD absorption lines (L3-0\,R0, L5-0\,R0, L7-0\,R0, L8-0\,R0 and W0-0\,R0) are clearly detected
and were fitted simultaneously. 
%The strongest constraint on the variation of $\mu$
%comes from the optically thin L3-0\,R0 transition. A consistent fit was performed first fixing the column 
%densities and Doppler parameters to the values from the fit assuming $\Delta\alpha$~=~0 and then relaxing 
%at once the wavelength of each component of each absorption line.
%these conditions.
The 3$\sigma$ limit reached here is $\Delta \mu/\mu < 9\times10^{-5}$.

Although the number of available lines and the signal-to-noise ratio do not allow to reach the level 
of accuracy achieved with H$_2$, it is important to pursue in 
this direction and to measure $\Delta \mu$/$\mu$ also from HD lines. This is very important given the
scarcity of possible independent measurements.

\subsection{NH$_3$}
\vskip -0.2truecm
Recently, Flambaum \& Kozlov (2007) showed that the high sensitivity of the NH$_3$ inversion transitions 
to a change in $\mu$ could be used to constrain the variations of this constant.
%by comparing the redshift of these transitions to those of transitions from other molecules. 
The only intermediate redshift system where this molecule is 
detected is the $z$~=~0.685 lens toward B~0218+357 (Combes \& Wiklind 1995, Henkel et al. 2005). They 
obtain a 3$\sigma$ upper limit on the variation of $\mu$ at this redshift of 6$\times$10$^{-6}$. 
Murphy et al. (2008) refined this limit to 2$\times$10$^{-6}$.
This technique has been applied by Levshakov et al. (2008) to NH$_3$ and other nitrogen rich 
molecules observed in our Galaxy with a sensitivity reaching 10$^{-7}$$-$10$^{-8}$.

\section{Combinations of constants}
\vskip -0.2truecm
\subsection{The 21~cm absorbers}\label{Sec3}
\noindent \vskip -0.2truecm
As the energy of the hyperfine H~{\sc i} 21-cm transition is proportional to 
the combination of three fundamental constants, $x=\alpha^2 g_{\rm p}/ \mu$, 
high resolution optical and 21-cm spectra can be used together to probe the combined
cosmological variation of these constants (Tubbs \& Wolfe 1980).
In the definition of $x$, $\alpha$ is the fine-structure constant, $\mu$ is the
proton-to-electron mass ratio and $g_{\rm p}$ is the gyromagnetic factor (dimensionless)
of the proton (see e.g. Tzanavaris et al. 2005). 
%(see e.g. Tzanavaris et al. 2005).

To apply this technique, the redshift of the 21~cm line must be compared
to that of UV lines of Si~{\sc ii}, Fe~{\sc ii} and/or Mg~{\sc ii}. Two difficulties
arise: (i) the radio and optical sources must coincide : as QSOs in the optical
can be considered as pointlike sources, it must be checked on VLBI maps that the corresponding
radio source is also pointlike which is not true for all quasars; (ii) the gas at the origin of the
21~cm and UV absorptions must be co-spatial: it is likely to be the case if the lines are narrow.
Therefore systems in which the measurement can be performed must be selected
carefully. Since the overall number of suitable systems is very small, they must be searched for. 

%Here again we are therefore facing  the problem that the systems where this test can be performed
%are very rare. 
For this reason we have embarked on a large survey to search for 21~cm absorbers
at intermediate and high redshifts. For this we first selected strong
Mg~{\sc ii} systems ($W_{\rm r}$~$>$~1~\AA) from the Sloan Digital Sky Survey in the redshift range suitable
for a follow-up with the Giant Meterwave Telescope (GMRT), 1.10~$<$~$z_{\rm abs}$~$<$~1.45.
We then cross-correlated the $\sim$3000 SDSS systems we found with the FIRST radio survey to
select the background sources having at least a S$_{\rm 1.4GHz}>50$\,mJy 
bright component coincident with the optical QSO. 
%with no extension on the radio image so that we are sure the background quasar
%is a radio point source at the $\sim$1~arcsec scale. 
There are only 63 sources fulfilling these criteria out of which we observed 35  
over $\sim$400~hours of GMRT observing time.

We detected 9 new 21~cm absorption systems. This is by far the largest number 
of 21-cm detections from any single survey. Prior to our survey no intervening 
21-cm system was known in the above redshift range and only one system was known 
in the redshift range $0.7\le z \le 1.5$. 
Our GMRT survey thus provides systems in a narrow redshift range where variations 
of $x$ can be constrained. For this, high resolution and high signal-to-noise ratio
observations of the absorbers must be performed to detect the UV absorption lines
that will provide the anchor to fix the exact redshift, the variations of $x$
being then constrained by the position of the 21~cm absorption line.
  
These UV observations exist for one of the system at $z_{\rm abs}$~=~1.3608
(see Fig.~\ref{21cm}). Although the UV data could be even better, a preliminary constraint
was obtained: $\Delta x/x$~$<$~10$^{-5}$ at the 3~$\sigma$ level.

 \begin{figure}
%\centerline{\includegraphics[width=9cm]{fig-name.pdf}}
%\centerline{\includegraphics[width=15cm]{21cm.jpg}} 
\centerline{\includegraphics[bb=8 127 574 703,width=13cm,clip=true]{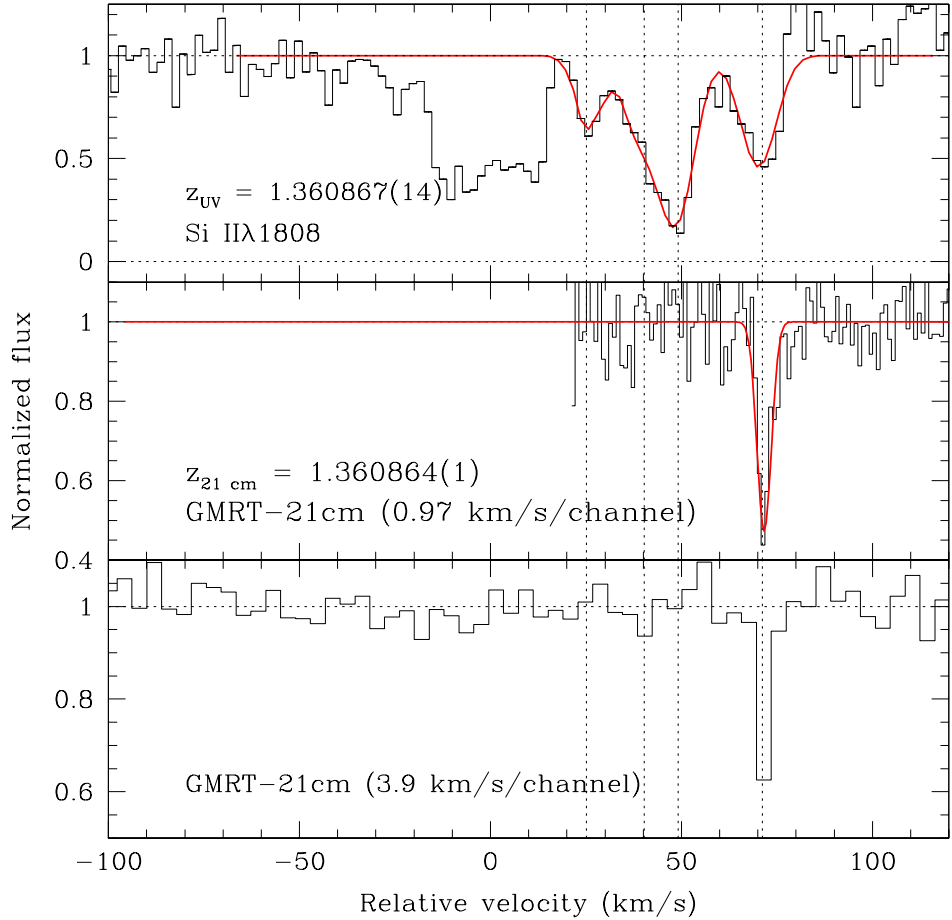}} 
\caption{Detection of 21~cm absorption in a cloud at $z_{\rm abs}$~=~1.3608.
The 21~cm component is very narrow and is associated with a UV component
well detached from the bulk of the Si~{\sc ii} absorption profile.
A comparison of the positions of these two transitions sets strong constraints
on the parameter $\alpha^2 g_{\rm p}/\mu$.
It is apparent that the errors will come from the determination of
the position of the UV Si~{\sc ii} absorption line.
}
\label{21cm}
\end{figure}

%\section{Other species}\label{Sec4}
 
\subsection{Other molecules: CO, OH, NH$_3$, HCO$^+$}
\vskip -0.2truecm
Other molecules can be used to derive strong constraints on fundamental constants.
Wiklind \& Combes (1999) noticed that a potential application of the observation of radio 
molecular absorption lines at high redshift is to check the invariance of constants. Radio lines
are well suited for this purpose because spectral resolution better than 1~km~s$^{-1}$ can
be achieved in the radio wavelength range.
% thatthe greatest frequency resolution can be achieved. 
By comparing the redshift of a molecular transition
to that of the 21~cm hyperfine H~{\sc i} line, it is possible to constrain a combination 
$\alpha^2 g_{\rm p}$($M_{\rm red}/m_{\rm p}$) of
the proton gyromagnetic ratio $g_{\rm p}$, $\alpha$ and the ratio of the reduced mass of the molecule
to the proton mass. They already put a 3$\sigma$ limit of 10$^{-5}$ 
on the variations of the above coefficient at $z_{\rm abs}$~=~0.25 and 0.68 using 
radio transitions from CO and HCO$^+$. A strong limitation of this technique is  that different
absorption lines may probe different volumes along the line of sight. This is also true
when comparing several CO lines as the opacity depends on the excitation conditions at
each point of the cloud. Note that this is also true for all techiques using absorption lines 
from different species. In addition, the systems where the test can be performed
are again, at the moment, very rare.

Similarly, beautiful observations of conjugate absorption and emission OH lines have been performed recently.
Kanekar et al. (2005) have detected the four 18~cm OH lines from the $z_{\rm abs}$~=~0.765 gravitational lens 
toward PMN~J0134$-$0931 with the 1612 and 1720~MHz lines in conjugate absorption and emission (see also Kanekar 
\& Chengalur 2004).
They compare the H~{\sc i} and OH absorption redshifts of the different components in this system and
also in the absorber arising from the $z$~=~0.685 lens toward B~0218+357 to place stringent constraints on 
changes in $F$~=~$g_{\rm p}(\alpha^2/\mu)^{1.57}$. They obtain $\Delta F/F$~$<$~4$\times$10$^{-5}$.

\section{Conclusion}\label{Conclusion}
\noindent \vskip -0.2truecm
Results are summarized in Table~1. Column \#1 gives the constant under study (see definitions in
the Text above), \#2 is for the method used, \#3 indicates the redshift at or the redshift range over which the 
measurement is performed, \#4 gives the constraints or measurement obtained and \#5 gives partial references (other 
references are given in the Text above). It is apparent that constraints reside in the 
range $\sim$0.3$-$1.5$\times$10$^{-5}$ at the 3$\sigma$ level over a redshift range of approximately
$0.5 - 2.5$. Note that Reinhold et al. (2006) do not claim detection. They are cautious enough to state 
that systematics dominate the measurements. Indeed, data and observed wavelength determinations are 
the same as those in Ivanchik et al. (2005). The only claimed detection of varying $\alpha$ is from
Murphy et al. (2003). It is therefore very important to increase the number of measurements 
and to improve the measurements themselves (e.g. Thompson et al. 2008)
to  reach the sensitivity level of at least $<$0.1$\times$10$^{-5}$ which should be possible in the
next few years. Future instrumentation on ELTs will undoutedly boost this
field by allowing to reach much better signal-to-noise ratios at higher spectral resolution (e.g. Liske et al. 2008). 
As discussed above, the strongest constraints may come from radio observations of molecules in 
the ISM of high redshift galaxies. A new area will be opened in this field by the
upcoming facilites such as ALMA and EVLA for high-redshift molecular studies and
SKA  pathfinders for 21~cm and OH surveys.   

\begin{table}%[!Ht]
\caption{\label{Tab1} Constraints on the comological variations of fundamental constants}  
\begin{tabular}{c c l c l}
\hline  
        & Method & Redshift & Constraint (10$^{-5}$)$^a$ & References$^b$ \\
\hline
$\alpha$ &    MMM       & $0.2 - 3.7$    & $-0.54\pm$0.12   & Murphy et al. (2003)      \\
         &              & $0.5 - 2.5$    & $<$0.30          & Srianand et al. (2004)    \\
         &    FeII      & 1.515          & $<$0.45          & Quast et al. (2004), Levshakov et al. (2005) \\  
         &    AD (SiIV) & $1.6 - 2.8$    & $<$1.3           & Chand et al. (2005)       \\
         &    AD ([OIII])& $0.15 - 0.8$  & $<$42 & Bahcall et al. (2004) \\
$\mu$    &    H$_2$     & 2.595, 3.025     & $<$2.1           & Ivanchik et al. (2005), Reinhold et al. (2006) \\
         &              & 2.595, 3.025     & $<$1.2           & King et al. (2008) \\
         &    HD        & 2.418          & $<$9            & Noterdaeme et al. (2008) \\
         &    NH$_3$    & 0.685     & $<$0.27           & Flambaum \& Kozlov (2007), Murphy et al. (2008) \\
$g_{\rm p}(\alpha^2/\mu)^{\nu}$ &  21cm & 1.361  & $<$1.0 & Srianand et al. (2009) \\
                                &    OH  & 0.685   & $<$4.0 & Kanekar et al. (2005) \\
                                & NH$_3$ & 0.685 & $<$5.0 & Flambaum \& Kozlov (2007) \\ 
                                & CO, HCO$^+$ & 0.25, 0.685 & $<$1.0 & Wiklind \& Combes (1997)\\ 
\hline
\multicolumn{4}{l}{$^a$ 3$\sigma$ for upper limits; $^b$ See Text for other references.}\\
%\multicolumn{4}{l}{$^b$ See Text for other references.}
\end{tabular}
\end{table}

{\acknowledgements
\footnotesize This work is based on observations collected during several
observing programmes at the European Southern Observatory with the 
Ultra-violet and Visible Echelle 
Spectrograph mounted on the 8.2~m KUEYEN telescope operated at the Paranal 
Observatory, Chile and at the Giant Meter-wave Radio Telescope in India. 
We gratefully acknowledge support from the Indo-French 
Centre for the Promotion of Advanced Research (Centre Franco-Indien pour 
la Promotion de la Recherche Avanc\'ee). PN was supported by an ESO PhD fellowship.
%We are indebted to the referees whose comments provided substantial input into this 
%paper without that the paper could not have been completed. We also thank the co-authors for their 
%negligible contributions. 
}

}
\end{article}

\end{document}